# An Image Processing Based Blur Reduction Technique in Smartphone-to-Smartphone Visible Light Communication System

**Vaigai Nayaki Yokar[1], Hoa Le-Minh[2], Zabih Ghassemlooy[3] and Wai Lok Woo[4]**

[1]Department of Mathematics, Physics and Electrical Engineering, Northumbria University, Newcastle Upon Tyne, United Kingdom
[2] Department of Mathematics, Physics and Electrical Engineering, Northumbria University, Newcastle Upon Tyne, United Kingdom
[3] Department of Mathematics, Physics and Electrical Engineering, Northumbria University, Newcastle Upon Tyne, United Kingdom
[4] Department of Computer and Information Sciences, Northumbria University, Newcastle Upon Tyne, United Kingdom

Corresponding author: Vaigai Nayaki Yokar (vaigai.yokar@northumbria.ac.uk).

**ABSTRACT** In this paper, we present a blur reduction technique for smartphone-to-smartphone visible light communications (S2SVLC). The key technique it to avoid the repeated scanning of the transmitted data and to lower the amount of data discarded at the receiver end of the S2SVLC system. This image processing method will improve the system recognition efficiency and data rate. The proposed method includes converting the red-green-blue (RGB) image into grayscale, applying contrast enhancement, scaling and binarizing the image to reduce the blur levels in the image. The experiment includes practical data acquisition and further processing and estimation in MATLAB. The experiment is carried out in different conditions like distance, rotation, and tilt also considering different surrounding illuminations like ambient light and no light conditions to estimate the blur levels in S2SVLC. In this experimental investigation two types of coding, American Standard code for information interchange (ASCII), and quick response (QR) code are used for data transmission in S2SVLC. The obtained results indicate that, the proposed technique is proven to improve the recovery efficiency to 96% in the receiver end at different conditions.

## I. INTRODUCTION

Visible light communication (VLC) has received significant attention recently [1] [2] [3]. VLC has grown continuously to potentially become an alternative to Radio Frequency (RF) technologies [4] [5] [6] . Unlike Wi-Fi and Bluetooth which uses RF based technology to transmit information, VLC transmits data via visible light. For sending data between smartphones, near field communication (NFC) is commonly used [7] as smartphone users rely on security and pre-existing RF technologies to transmit information. Alternatively, VLC can be used to setup a short link communication channel between two smartphones. This idea opens a new field in VLC known as smartphone-to-smartphone visible light communications (S2SVLC) [4][6].

With the increase in the field of optical camera communication (OCC) [4][6]. A screen to camera link-based communication emerges as a new method in device-to-device communications [7][8] . It includes liquid crystal display (LCD) and a camera on the smartphone to setup the S2SVLC communication link [4][5][6].

In S2SVLC system, the data is encoded as a quick response (QR) or American standard code for information interchange **(**ASCII) image and is transmitted from a transmitter (Tx) smartphone to the receiver (Rx) smartphone by establishing a screen to camera link [4][5]. The QR code is a 2-D Matrix code. QR makes use of four input modes i.e., numeric, alphanumeric, byte/binary and kanji/kana to store data [9]. QR codes can store maximum of 7089 numeric characters, 4296 alphanumeric characters, 2953 binary/byte characters and 1817 kanji/kana characters [10]. The ASCII code is the alternative data transmission method used in this paper. The ASCII is a 7-bit code. The eighth bit which is one full byte is traditionally used for checking purpose.

One of the main problems in S2SVLC based OCC is the image blurring especially due to growing distance, tilt or rotation angles between the Tx and the Rx smartphones. The other reason for blurring is the surrounding illumination levels and when the camera is in mobility [4][10]. Blurred image restoration is the hot topic in image processing [10].

In this paper we propose an image processing-based algorithm to help with the blurred image restoration in the S2SVLC. This proposed method is less complex when compared to the other methods.

## II. LITERATURE REVIEW

The groundwork of S2SVLC was laid by COBRA [11] followed by Rainbar [12], SoftLight [13], TETRIS [5], Surfing method and Flashing method [6]. The COBRA system [11] can achieve high speed barcode streaming between smartphones based on lightweight image processing technique. But it inputs system throughput by using highly customized barcodes, which are not widely adopted in practice. The main problem faced by cobra were the low resolution of the smartphone and small smartphone screen. Nevertheless, the issue with blurring is still one of the major challenges in S2SVLC with the improved resolution and screen size overtime [4][5]. The lightweight image processing developed in COBRA system are not efficient to the widely used barcodes like QR and ASCII. As a result, COBRA system can't be widely used in practice [4]. In COBRA system the display rate of color barcode frames on the screen is smaller than half of the capture rate of the camera, the receiver is likely to capture more than one image from the same frame with different degree of blur. It is a wasteful of time and resources to process captured images of the same frame. COBRA [11] used the HSV (hue-saturation-value) colour space to improve the quality of images as HSV is much more effective in recovering colors from the blurred images and therefore the accuracy of colour recognition, which however is very time-consuming. For example, it takes 16 ms for COBRA to decode a captured image while 12 ms is used for HSV colour space, which severely affects the processing performance for barcode decoding [12]. The challenges of color barcode streaming in RAINBAR system [12] are dynamic environments, lens distortion, high display rate, angle, and distance between the screen and camera. The same blur assessment method proposed in COBRA system is used in RAINBAR. Instead of RGB, SoftLight [13] modulates data in YUV color space. RGB color space is composed of three components (Red, Green, Blue). All three components consist of brightness ingredient which are sensitive to ambient light and might easily leak to neighboring pixels due to blur effect. The higher frame rate in SoftLight means higher bandwidth and higher blur effect due to rolling shutter. Brightness is extracted as an independent component Y, and the other two chrominance components U and V are also independent. Such high false-negative prohibits the use of derived soft hints to establish an effective channel. The inaccurate estimation of soft hint is mainly due to the inherent intra-frame colour interferences of current VLC colour modulation [13].

In TETRIS [5], the low ambient light during testing can also cause blurring between color blocks. It is caused by misclassification of points between the border of two colors, where blurring can skew the colors. TETRIS doesn't support adaptive blur awareness. The accuracy of TETRIS could be improved with additional blur awareness. The most time-consuming processing components are to find the timing reference strips and determine the colour of all blocks using the grid, which is a common problem in TETRIS [5]. In the flashing and surfing method [6], the HSV color space is carried out for image processing. As mentioned, HSV is very time consuming. Also, in HSV there is an undefined scenario for hue and for saturation, an issue arises when the maximum value of RGB is 0 (Black color).

There are two types of image restoration methods namely non-blind restoration method and blind restoration method. Non-blind restoration methods are least mean square error (LMSE), wiener filtering and inverse filtering. While blind restoration methods are neural network and super-resolution. The receiver needs to know the mean value and variance of noise for LMSE and wiener-based restoration methods which is not feasible for S2SVLC [14]. In Inverse filtering, the image information will be covered by increased noise levels in case of low SNR value [15]. While the blind restoration methods are also not suitable for S2SVLC system due to high computational complexity and reduce the achievable data rate of the system [16].

Here, we propose a novel blur reduction image processing algorithm for S2SVLC for improved accuracy and system performance. The image processing model is used in the receiver end to improve the recovery efficiency and data rate of the system.

## III. PRINCIPLE

The schematic diagram of the S2SVLC system is represented in Fig. 1. In S2SVLC, a stream of data (text of any other media format) is converted into binary stream of data (0's and 1's), which are then encoded into and image where 0's and 1's is assigned with a white and black color pixel respectively. This process converts binary stream of data into image. The captured image is proposed using the image processing algorithms in the Rx end and then is decoded back to text.

In this work, we propose a S2SVLC image processing algorithm which aims to solve the main problem of image blurring especially due to growing distance, tilt or rotation angles between the Tx and the Rx smartphone. The proposed S2SVLC system includes an image processing technique which allows loading the Rx image into Matlab, converting the RGB image to grayscale image, applying contrast enhancement, scale and binarize the image. The image processing improves the recovering efficiency and data rate of the S2SVLC communication system. The proposed image algorithm modulates the data in grayscale which improves the sensitivity to ambient light. The S2SVLC system is represented in the Fig.2.

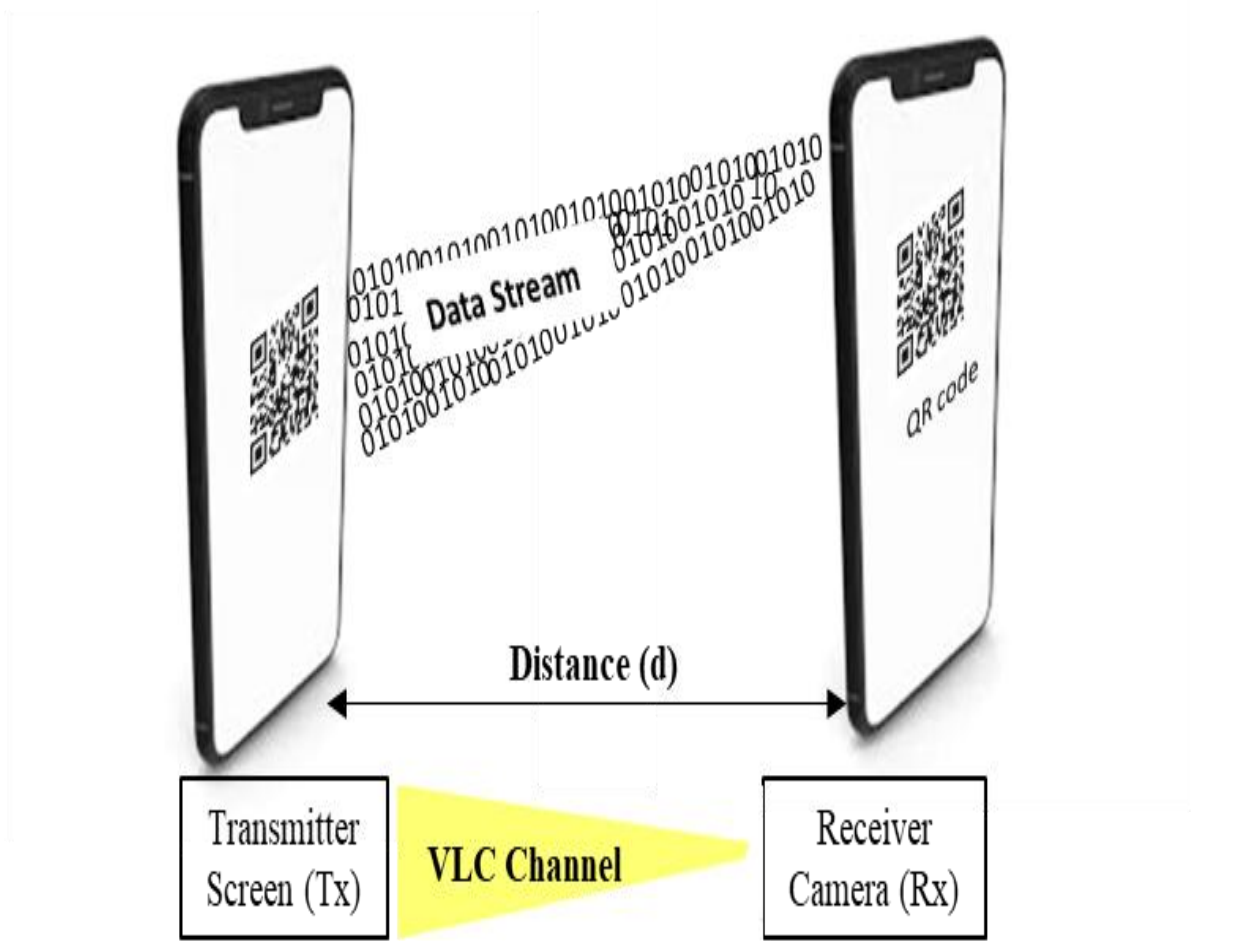

**FIGURE 1.** Representation of the system.

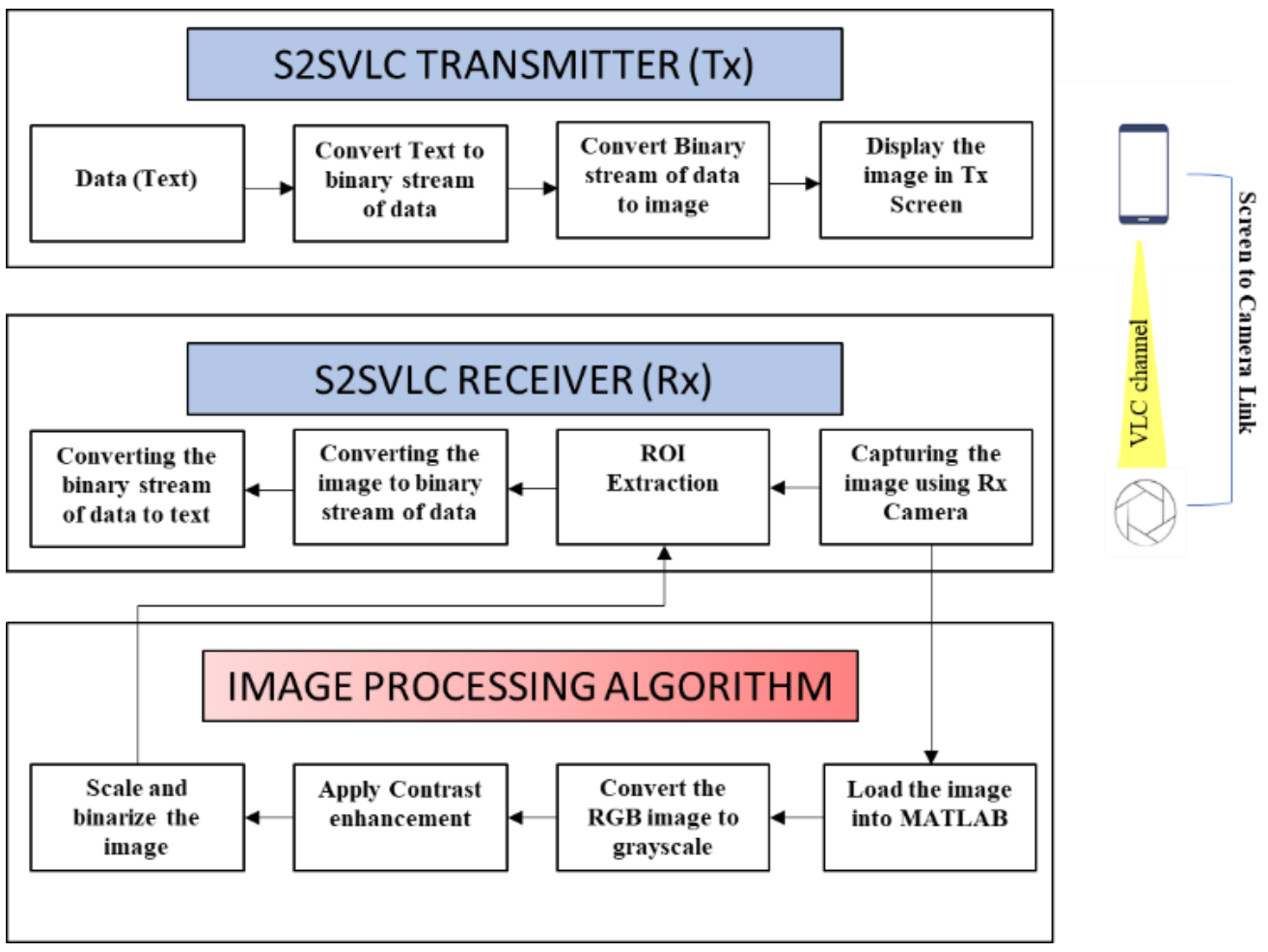

**FIGURE 2. System block diagram of the S2SVLC system with the proposed algorithm.**

The proposed algorithm improves the blurring effect of images in SVLC communication system, which has advantages of low complexity. The details of the sub system of the proposed model in Fig 3. is represented as follows:

### *A. SMARTPHONE SCREEN*

The Google pixel 6 Pro smartphone screen is used as an optical transmitter. As proposed by IEEE 802.15.7 standard for optical wireless communication it is common to find optical transmitters confronted by phosphorescent light emitting diode (LED) [19]. The google pixel 6 pro has a 170 mm full-screen display with the aspect ratio of 19:5:9. The display has a HDR support with > 1,000,000:1. Note that the typical refresh rate of the screen is 120 Hz. The smartphone screen is based on RGB888 color scheme, which is made of three prime colors red, green, and blue [20]. Each color has 256 possible levels hence it can be encoded as an 8-bit code word. The combination of all the prime colors will generate more than 16.7 million ($2^{24}$) possible colors on the smartphone screen [21].

### *B. SMARTPHONE CAMERA*

The CMOS based camera is cheaper and more power efficient than the CCD based image sensing technology [22]. The camera subsystem includes the implementation of fundamental image processing algorithms that converts the raw Bayer output from the camera sensor to a full-fledged image that can be consumed by applications and users. we intuitively assume, the output of the camera sensor only contains information on one color per pixel as compared to three colors (RGB) per pixel. The frame rate of the camera is around 60 fps.

### *C. SMARTPHONE VLC SYSTEM*

As seen in Fig. 1, a typical on-off keying (OOK)-based S2SVLC system consists of two devices that are utilized as a Tx and a Rx for displaying data frames and collecting frames, respectively. The data stream is divided into a number of data frames at the Tx after being transformed into a binary bit format (i.e., an ASCII character in byte form). Each frame is made up of an image that has been divided into $M \times N$ cells. In OOK or systems that use colour shift keying (CSK), these cells are colored white for '0's and black for '1's. Following data populating frames, they are displayed in a predefined order on the Tx's screen. At the Rx, a camera is employed to extract data frame by frame after identifying the region of interest (ROI) in the images that were taken. To accomplish ROI detection, each frame is made up of crucial information such as the marking area, sequential number, error correction, and timestamp in addition to the data payload. A frame is split into MXN cells after a successful detection, and then these cells are quantized, converted back to a binary stream, and then returned to their original data format.

## IV. THEORITICAL MODELLING

The OOK is the most reported modulation technique in S2SVLC due to its simplicity [4][5].A bit ‘1’ is simply represented as an optical pulse that occupies the entire part of the bit duration while a bit ‘0’ is represented by the absence of an optical pulse. A non-return-to-zero (NRZ) scheme is applied is applied to the S2SVLC system. In NRZ, a pulse with duration equal to the bit duration is transmitted while in return-to-zero (RZ) the pulse occupies only the partial duration of the bit. Hence, the envelop for OOK-NRZ is given by (1):

$$p(t) = \begin{cases} 2\,P_r & for\ t\ \in [0, T_b] \\ 0 & elsewhere \end{cases}, \qquad (1)$$

where, $P_r$ is the average power and $T_b$ is the bit duration.

The simplicity of the OOK has led to its use in optical communication systems operating below 4 Mbps [23] and the S2SVLC data rate is 324 kbps. CSK is a visible light intensity modulation scheme, outlined in IEEE 802.15.7 [24], that transmits data imperceptibly through the variation of color emitted by red, green, and blue color channels. Using CSK, the bits are assigned with colors in S2SVLC. On the receiver side, the image sensors receive the color symbols in the form of different color bands in a frame. Depending on the Rx hardware and the number of colors the sensors can capture, higher CSK modulation can be implemented to dramatically improve the data rate [25].

## V. PROPOSED BLUR REDUCTION ALGORITHM

The blur reduction method includes the following processing steps:

| **BLUR REDUCTION TECHNIQUE** | |
|---|---|
| **Step 1** | Import a RGB encoded image and convert into a grayscale image (using the conversion equation. |
| **Step 2** | Apply the contrast enhancement to the converted grayscale image to improve its visual quality. |
| **Step 3** | Normalize the minimum and maximum values of the grayscale image to ensure that a correct detection threshold level is being used. |
| **Step 4** | Binarize the pixels based on the obtained threshold (in step 3) to restore the black and white image. |

### *A. STEP 1 – CONVERT RGB IMAGE INTO GRAYSCALE*

The received blurred image from the Rx camera as represented in Fig. 2. was imported into MATLAB script and was found to have bit errors. The bit errors can be due to superimposition of the neighboring pixels, brightness ingredient leaking into the neighboring pixels due to ambient light sensitivity, motion blur, brightness level of transmitter screen, channel noise and more [4]. The RGB (red-green-blue) has three color channels: red channel, green channel, and blue channel. However, grayscale has only one channel. We convert the RGB image into grayscale (*i*) to store a single-color pixel of RGB, we need 24 bits (8 bit for each color component) but when we convert RGB to grayscale image, only 8 bit is required to store single pixel value of the image [26], (*ii*) grayscale representations are often used for extracting descriptors instead of operating on color images directly as grayscale simplifies the algorithm and reduces the computational requirements [27]. The RGB to grayscale conversion is represented in (2):

$$\text{Grayscale} = \frac{R+G+B}{3}, \quad (2)$$

Where, *R*, *G* and *B* stands for red, green, and blue channel respectively. Theoretically, the expression (2) is accurate but while writing a code for conversion, the 8-bit unsigned integer (uint8) overflow error may occur if the sum of *R*, *G*, and *B* is greater than 255. To prevent this issue, *R*, *G*, and *B* should all be calculated separately to get the grayscale [4], as follows:

$$\text{Grayscale} = \frac{R}{3} + \frac{G}{3} + \frac{B}{3}. \quad (3)$$

If, we need to convert weighted RGB into grayscale the expression (4) allows for chroma subsampling, because human vision has finer spatial sensitivity to luminance difference than chromatic differences, which allows the system to store and transmit information at lower resolution.

$$\text{Grayscale} = 0.299R + 0.587G + 0.114B. \quad (4)$$

Grayscale images are (i) very common, in part because much of today's display and image capture hardware can only support 8-bit images; and (ii) entirely sufficient for many tasks and so there is no need to use more complicated and harder-to-process colour images.

### B. STEP 2 – APPLY CONTRAST ENHANCEMENT

The contrast enhancement technique transforms the intensity values of an input image to improve the quality of the visual image. By increasing the brightness difference between items and their backgrounds, contrast enhancements increase the visibility of objects in the scene. Although these might both be done in one step, contrast enhancements are commonly conducted as a contrast stretch followed by a tonal enhancement. While tonal enhancements boost the brightness differences in the shadow (dark), midtone (grey), or highlight (bright) regions at the expense of the brightness differences in the other regions, a contrast stretch improves the brightness differences equally across the dynamic range of the image. The contrast enhancement function increases the contrast of the image by mapping values of the input intensity image to new values such that, by default, 1 % of the data is saturated at low and high intensities of the input data [28]. The field of view of the system increases progressively using the contrast enhancement technique [4]. In (4) , $P_{in}$ is the original pixel value and $P_{out}$ is the derived output pixel value. Let's assume that (i) *a* and *b* are the lower and upper limits of pixel values (i.e., 0 and 255, respectively); and (ii) *c* and *d* is the existing lowest and highest pixel values in the captured image, respectively. Thus, each pixel can be scaled as [29]:

$$P_{Out} = (P_{in} - C)\left(\frac{b - a}{d - c}\right) + a, \quad (4)$$

Also, the list of figure properties such as image size, image type and image position are set to a constant value before applying the further techniques.

### C. STEP 3 & 4 – SCALING AND BINARIZING THE IMAGE

Image scaling is the process that increases or decreases the size of a digital image. The simplest method is copying every original pixel from the source image to its matching spot in the larger image. In the larger image, this will result in gaps between the pixels that are filled by giving the vacant pixels the colour of the source pixel that is to the left of the current location. In essence, this increases the size of an image and its data. While this approach, known as nearest neighbor, is successful at reducing data loss but the quality of the final image usually degrades because of image scaling since the expanded blocks of individual pixels will be obvious. Binarization is the process of converting grayscale image into a black and white image (i.e., 0 and 255 pixels respectively). This can be achieved with the help of a process called thresholding. We can get binary images using the process called thresholding [4]. We use adaptive thresholding as the image has been captured at different lighting conditions. In case of adaptive thresholding, the algorithm divides the image into smaller regions and automatically determines the threshold values for small regions. To verify the scaling process further the starting low threshold, point and starting high threshold points should be derived from the gray levels [4] . Then the images are scaled to the threshold level. The images are converted into a binary data stream using the derived threshold values.

## VI. EXPERIMENTAL INVESTIGATION, RESULTS AND ANALYSIS

The system includes two smartphone (Google Pixel 6 Pro) used as the Tx and Rx which are placed on the same axis as represented in Fig 4.

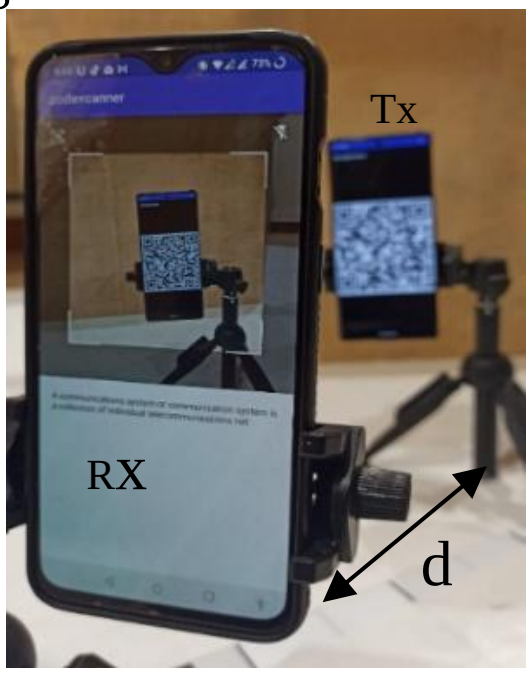

**FIGURE 3. Real time system representation**

Using the experimental testbed as represented in Fig 4, the encoded images were captured for different transmission ranges (between 0 to 50 cm), tilt angle (-50 to 50 degrees) and rotation angle (50 to -50 degrees). The bit-error-rate (BER) in the received image was investigated at different transmission environments like ambient lighting, and no light conditions. The ambient lighting describes the even lighting conditions and for the no light the image transmission was carried out in a dark room. The key parameters of the system are represented in Table I.

TABLE I.
KEY SYSTEM PARAMETERS

| **Parameter** | **Value** |
|---|---|
| No. of camera pixels | 50 MP |
| Aperture | f/1.85 |
| Sensor type | CMOS, Laser detection autofocus (LDAF) |
| Sensor model | Sony IMX386 |
| Sensor size | 1/1.31" wide |
| Camera frame rate | 60 fps |
| Camera focus | Auto focus |
| Tx display | OLED |
| Tx display size | 6.41 inch |
| Tx frame rate | 60 fps |
| No. of binary bits transmitted | 40,000 |
| Size of the image (cell x cell) | 200 x 200 |
| Distance between the Tx and the Rx (*d*) for experimental investigation | 0-50 cm |
| Tilt angles of the Rx with respect to the Tx for experimental investigation | -50 – 50 degrees |
| Rotation angles of the Rx with respect to the Tx for experimental investigation | -50 t0 50 degrees |

The carried out experimental investigation has demonstrated an important step to realizing a S2SVLC in real scenario. In the first experimental investigation, we investigated the BER improvement with proposed algorithm and the conventional system (i.e., without image processing) under different lighting conditions over a link span of 0 - 50 cm, see Fig. 4.

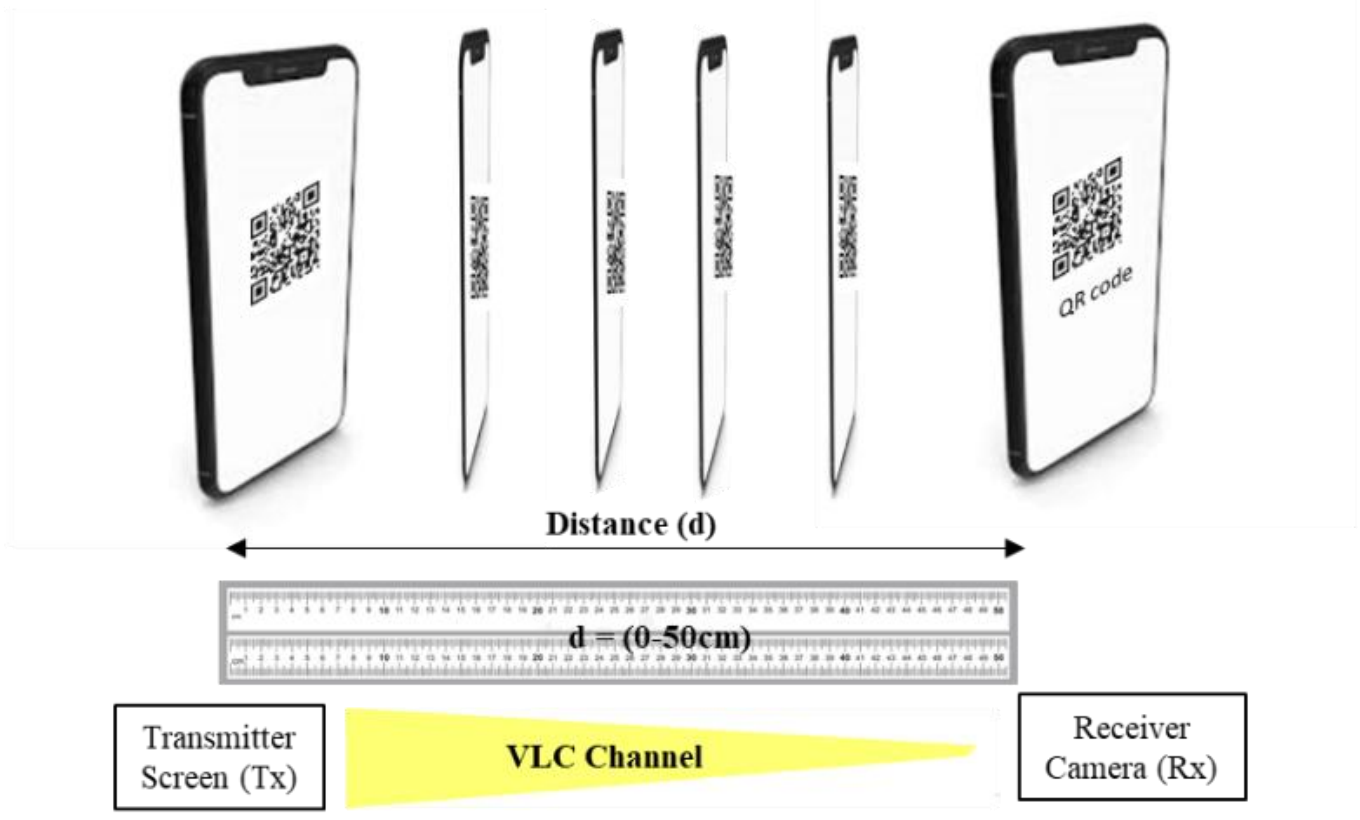

**FIGURE 4. Schematic representation of measurement at different distances in S2SVLC.**

The images are captured in both ambient light (123 lumens) and no light conditions. About 40,000 bits are transmitted and the BER is calculated over the link span as represented in Fig. 5. The BER is noticed to be lower in ambient light conditions compared to the no light conditions, as the amount of light entering the camera in no light conditions causes superimposition of the pixel which adds blur to the images. We conclude, the improved results after image processing improves the BER in both ambient and no light surrounding conditions. The surrounding illuminance plays an important role in the S2SVLC system. Over the linkspan (0-50) cms it is noticed that the BER increases with distance in both ambient and dark light condition. The BER increases as the distance between the Tx and Rx increases which induces channel noise in the S2SVLC system.

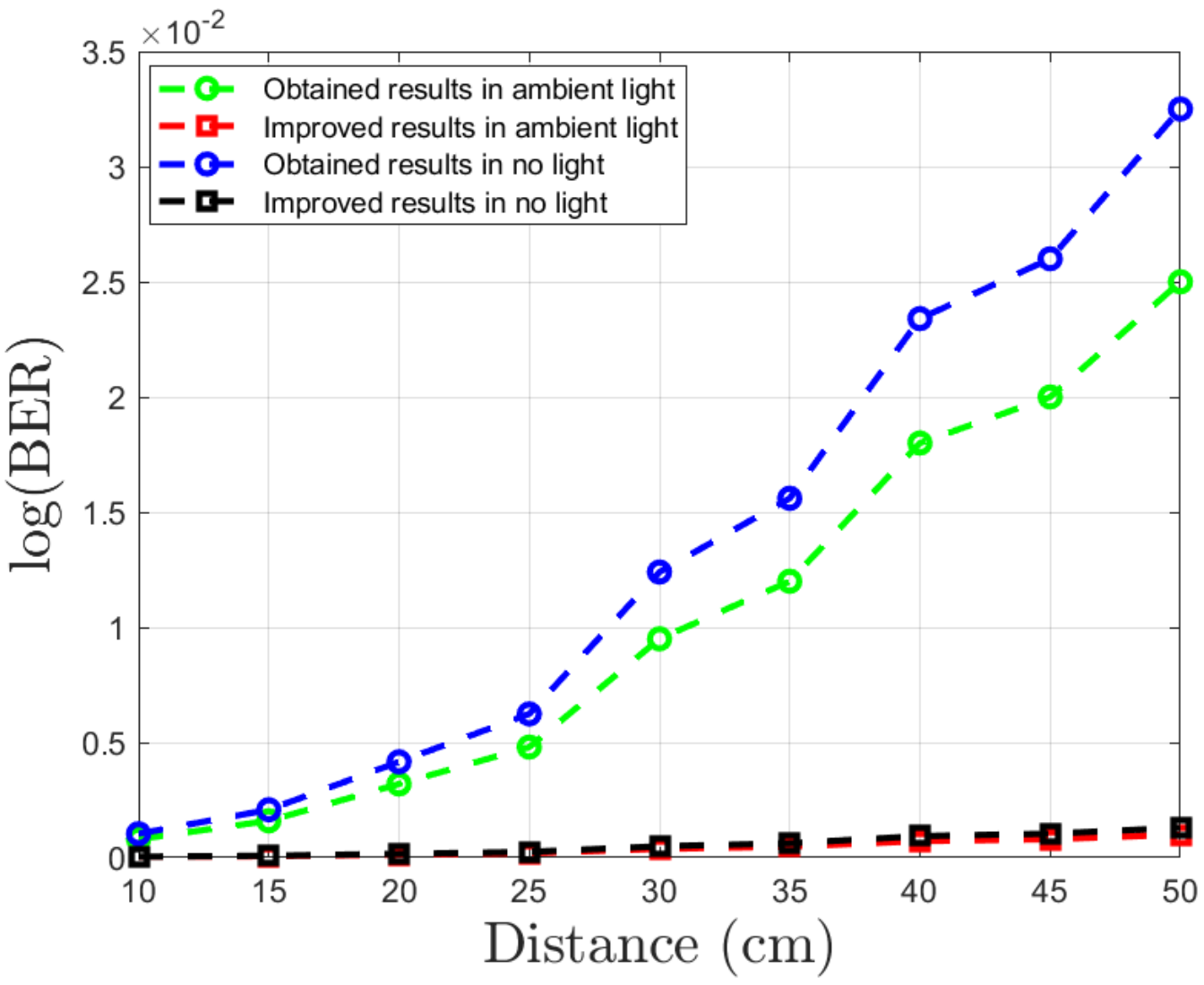

**FIGURE 5. BER vs Distance(cm) in S2SVLC.**

In the second experimental investigation, we investigated the BER using the proposed method and the conventional scheme under different lighting conditions (i.e., ambient light and no light) surrounding luminance at 15 cm with the span of tilt angles varying from -50 to 50 degrees. The experimental setup schematic is represented in Fig. 6.

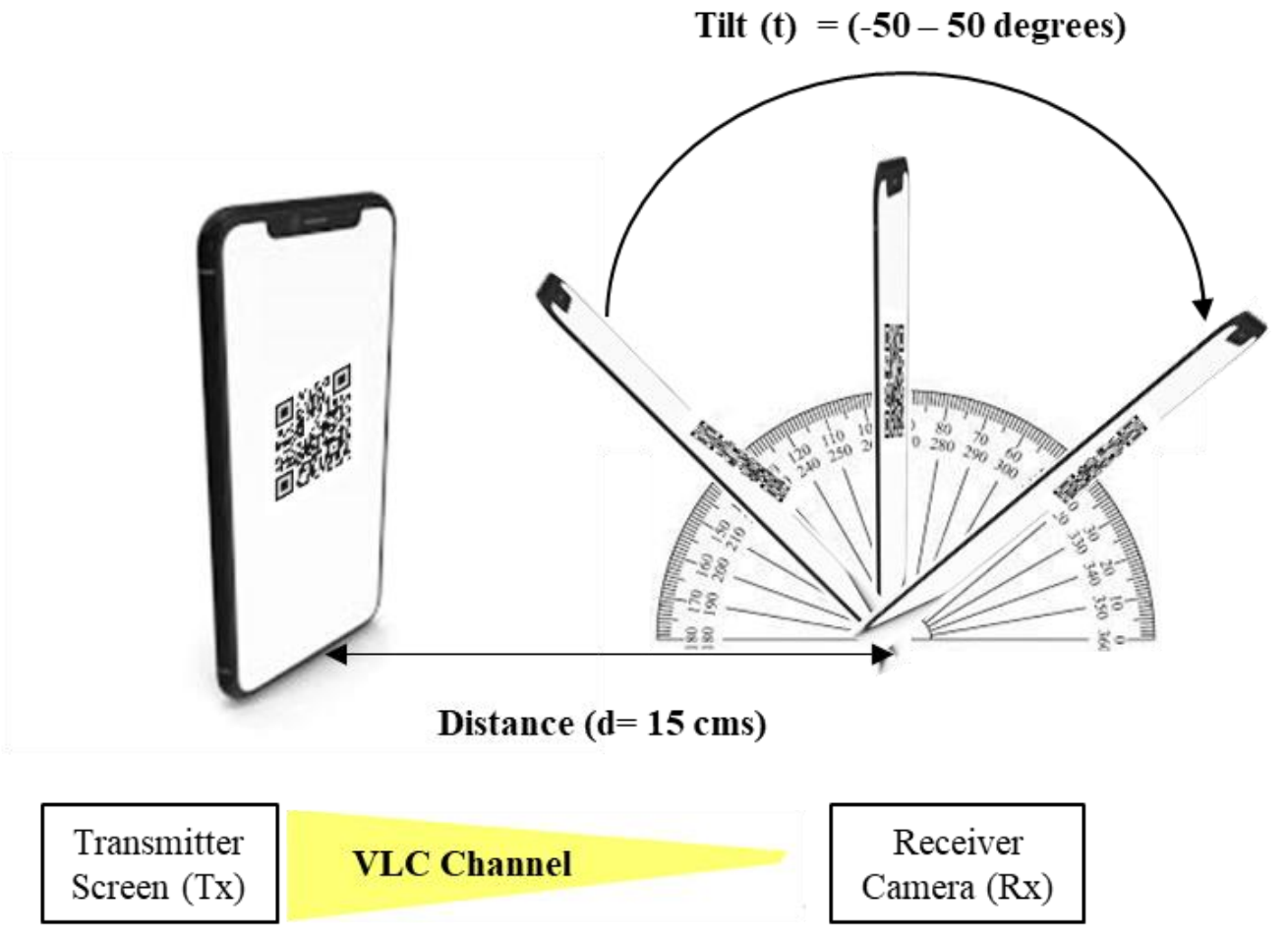

**FIGURE 6. Schematic representation of measurement at different tilt angles in S2SVLC.**

The improved results are represented in Fig. 7. The BER is noticed to be lower when the Tx is in no tilt position. With the increase in the tilt angles, there is an increase in the BER. The BER increases with the increasing tilt angles as the amount of light entering the receiver camera drastically reduces as the receiver is not in line of sight. After the span of (-50,50) degrees the field of view (FOV) of the received image is Rx smartphone drastically reduces, which makes it impossible to process the data as the number of bits recovered in the receiving end is very low. The ambient surrounding illuminance provides better results when compared to the no light condition in the experimental investigation. The ambient light induces BER as the RGB image is sensitive to ambient light which results in the brightness ingredient leaking into the neighbouring pixels. The bit error rate reduces to around 0.385 in both -50 and 50 tilt angles. So, the images are not captured for more than the specified link span (-50 - 50) degrees.

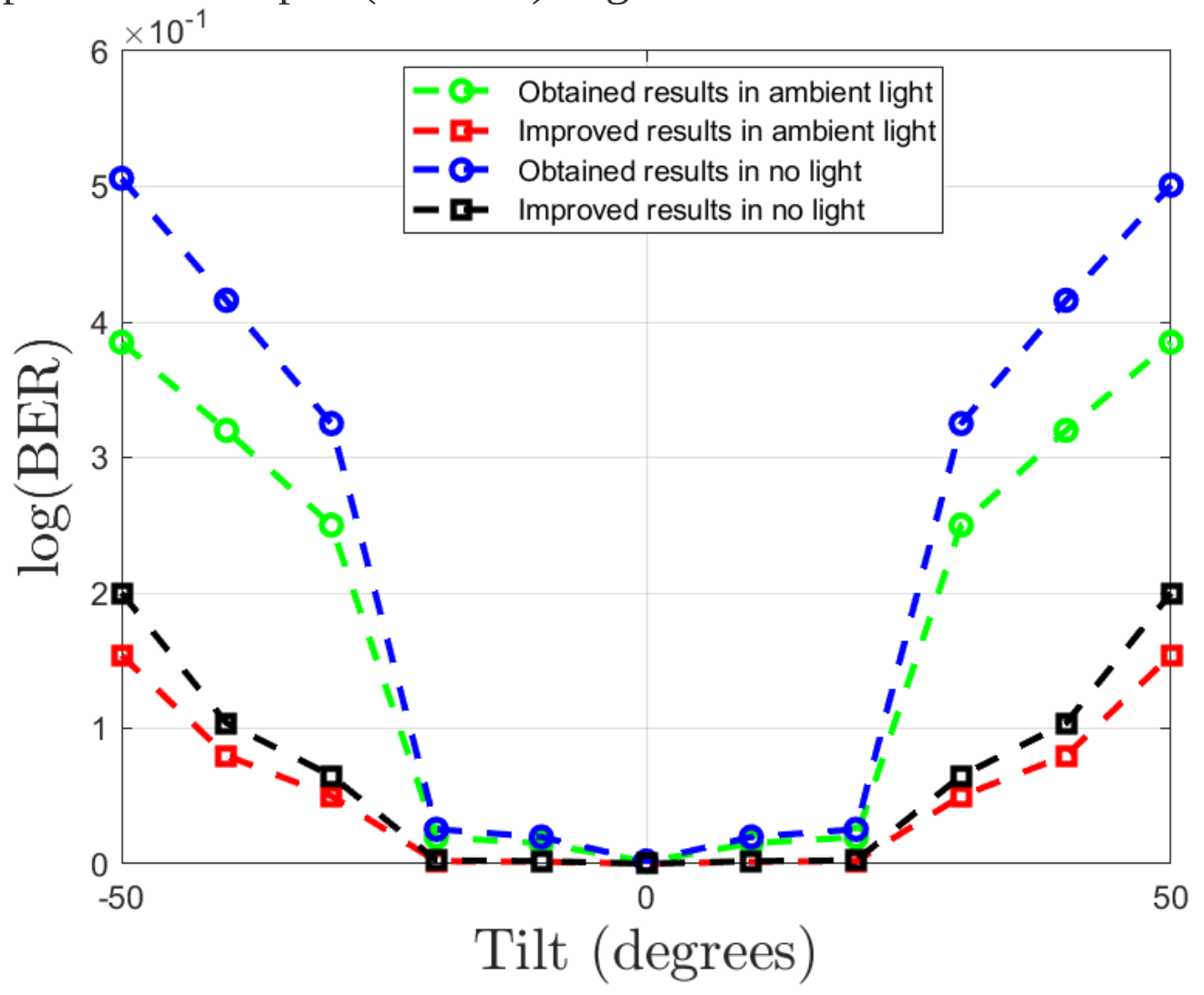

**FIGURE 7.** **BER vs Tilt (degrees) in S2SVLC.**

In the third experiment, we investigated the BER using the proposed method and the conventional scheme under different lighting conditions at 15 cm with the span of rotation angles of -40 to 40 degrees, see Fig. 8. The images captured in S2SVLC with varying rotation angles at 20cm distance are processed using the proposed algorithm. The improved results are represented in Fig. 9.

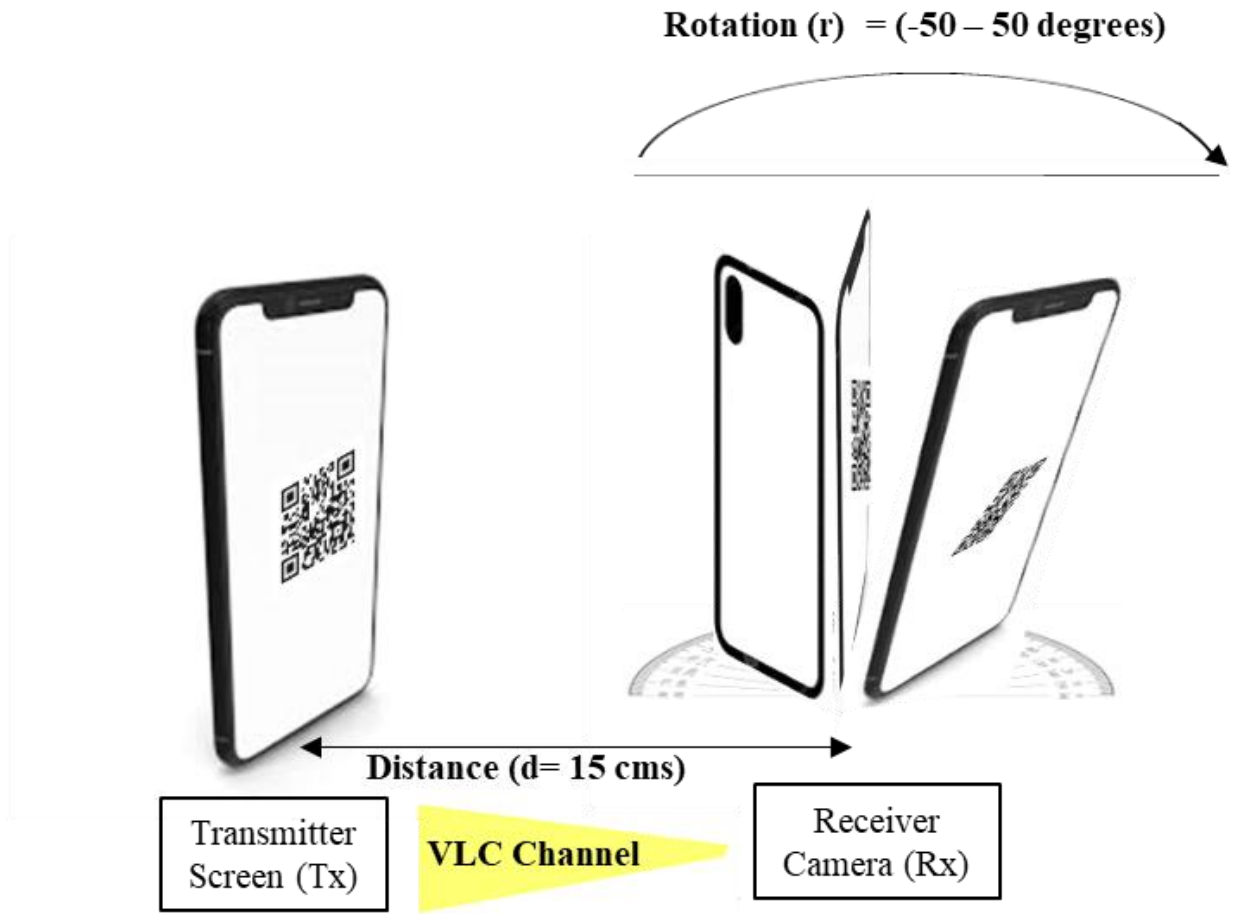

**FIGURE 8.** **Schematic representation of measurement at different rotation angles in S2SVLC.**

The BER is noticed to be lower when the Tx is in direct line of sight with the Rx. With the increase in the rotation angles, there is an increased BER in the S2SVLC system. The BER increases with the increase in rotation angles, as the amount of light increasing into the Rx camera reduces. After the span of (-50,50) degrees the FOV of the received image is Rx smartphone drastically reduces, which makes it impossible to process the data as the number of bits recovered in the receiving end is very low. The bit error rate reduces to around 0.35 in both -50 and 50 rotation angles. So, the images are not captured for more than the specified link span (-50 - 50) degrees. The field of view of capturing the entire frame also reduces after the span angles. The system is barely guessing the encoded information with the increased BER, the screen to camera link is lost.

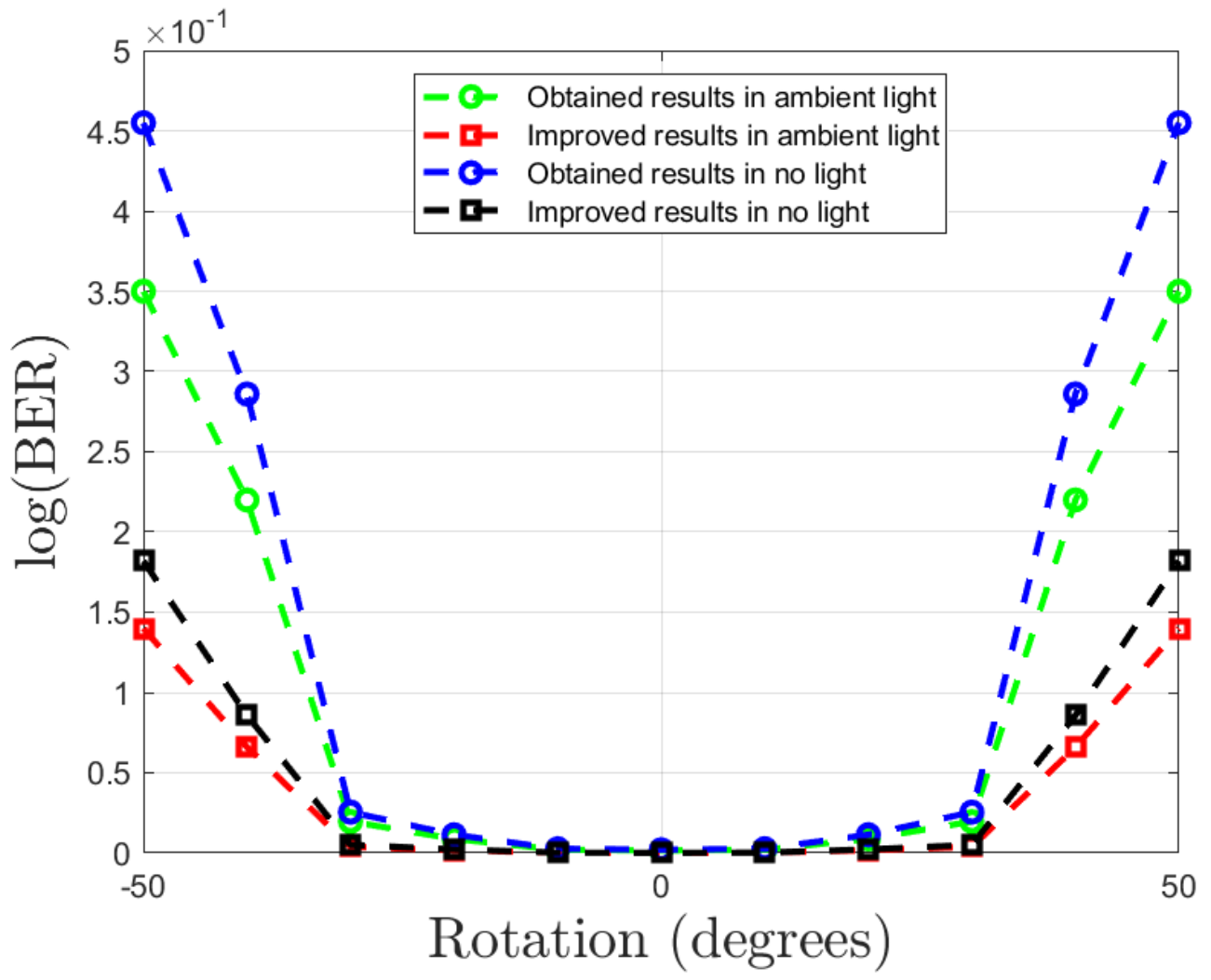

**FIGURE 9.** **BER vs Rotation (degrees) in S2SVLC.**

The BER increases with the increasing distance, tilt, and rotation angles between the Tx and Rx in S2SVLC. The BER increases due to various factors like superimposition, blurring, mobility, distance, tilt, rotation angles, Tx screen brightness and surrounding illumination. The BER can be very high when the received image is blurred due to mobility. The image processing algorithm has improved the detection and processing of the data in S2SVLC by 96%. The S2SVLC system with proposed algorithm achieves a higher data rate than the conventional system. The proposed algorithm increases the efficiency of the system by processing blurred images rather than discarding them in the conventional system.

## VII. CONCLUSION

In testing our algorithm, we focused on optimizing our results by introducing different surrounding illuminance, distance, tilt, and rotation span, while maintaining accuracy of the system. The images were enhanced using the proposed algorithm. The proposed system obtains higher results in both ambient and no light surrounding illuminance conditions. Also, the proposed algorithm improves the S2SVLC system performance by 96%. The blur awareness in S2SVLC is highly focused in this paper and the proposed method is also found to be less complex when compared to the other existing method discussed in the paper.